\documentclass[%
aip,
amsmath,amssymb,
reprint
]{revtex4-1}

\usepackage{graphicx}% Include figure files
\usepackage{dcolumn}% Align table columns on decimal point
\usepackage{bm}% bold math
\usepackage[mathlines]{lineno}% Enable numbering of text and display math
\usepackage[utf8]{inputenc}
\usepackage[T1]{fontenc}
\usepackage{mathptmx}

\usepackage{hyperref}% add hypertext capabilities
\usepackage{amsmath}
\usepackage{amssymb}
\usepackage{wasysym}
\usepackage{mathtools}
\usepackage{physics}
\usepackage{boxedminipage}
\usepackage[title]{appendix}
\usepackage{subfigure} % subfig
\usepackage{xcolor}

\begin{document}
	
	\preprint{AIP/123-QED}
	\title{Spin-memory loss induced by bulk spin-orbit coupling at ferromagnet/heavy-metal interfaces}
	
	\author{Mijin Lim}
	\author{Hyun-Woo Lee}%
	\email{hwl@postech.ac.kr}
	\affiliation{
		Department of Physics, Pohang University of Science and Technology, Pohang, 37673, Republic of Korea
		%\\This line break forced with \textbackslash\textbackslash
	}%

	\date{\today}% It is always \today, today,
	
	\begin{abstract}
		A spin current through a ferromagnet/heavy-metal interface may shrink due to the spin-flip at the interface, resulting in the spin-memory loss. Here we propose a mechanism of the spin-memory loss. In contrast to other mechanisms based on interfacial spin-orbit coupling, our mechanism is based on the bulk spin-orbit coupling in a heavy-metal. We demonstrate that the bulk spin-orbit coupling induces the entanglement between the spin and orbital degrees of freedom and this spin-orbital entanglement can give rise to sizable spin-flip at the interface even when the interfacial spin-orbit coupling is weak. Our mechanism emphasizes crucial roles of the atomic orbital degree of freedom and induces the strong spin-memory loss near band crossing points between bands of different orbital characters.
	\end{abstract}
	
	\maketitle
	
	A spin current can be both generated~\cite{Miron2011,Liu2012} and relaxed~\cite{Nguyen2014} by the spin-orbit coupling (SOC). Accurate determination of spin current characteristics such as the spin generation efficiency (spin Hall angle) and the relaxation rate (or spin diffusion length) is difficult, however. Reported experimental values of these characteristic parameters are spread sometimes over one order of magnitude even for seemingly same systems~\cite{Liu2011}. An important source of this spread is ferromagnet (FM)/heavy-metal (HM) interfaces~\cite{Kurt2002,Nguyen2014,Zhang2015}. Experimental schemes to quantify a spin current, such as spin-orbit torque measurements~\cite{Miron2011,Liu2012} and spin pumping measurements~\cite{Saitoh2006,Ando2011}, utilize a FM/HM bilayer and the spin polarization can be reduced at the FM/HM interface as a spin current passes through it. It is thus important to take account of this so-called spin-memory loss~\cite{Kurt2002,Bass2007spin,Rojas-Sanchez2014,Nguyen2014} for quantitative analysis of a spin current~\cite{Liu2014PRL,Nguyen2016PRL,Berger2018,Swindells2019,Zhu2019PRL_SML_TMS}. 
	
	There are ongoing effors to clarify mechanisms of the spin-memory loss~\cite{Chen2015a,Chen2015,Amin2016,Borge2017,Dolui2017,Tokac2015,Belashchenko2016,Gupta2020,Tao2018,Zhu2019spin}. Many theoretical~\cite{Chen2015a,Chen2015,Amin2016,Borge2017} and experimental~\cite{Tao2018,Zhu2019spin} works focused on effects of the {\it interfacial} SOC such as the Rashba spin-momentum coupling. However much less attention has been paid to roles of the {\it bulk} SOC (within a HM) despite the fact that the bulk SOC in a HM is strong in most experimental situations with the strong spin-memory loss. The bulk SOC effects are taken into account in a few first-principles calculations~\cite{Tokac2015,Belashchenko2016,Gupta2020} but are not analyzed explicitly.
	
	In this study, we investigate effects of the bulk SOC $\sim \mathbf{S}\cdot \mathbf{L}$ [Eq.~\eqref{eq:H_SOC}] on the spin memory loss at a FM/HM interface. In centrosymmetric systems such as fcc Pt bulk, the bulk SOC does not induce any spin splitting. Thus in order to investigate effects of the bulk SOC, it is crucial to take into account the atomic orbital degree of freedom. We use a model Hamiltonian [Eq.~\eqref{eq:H_tot}] that takes into account the atomic orbital degree of freedom explicitly and demonstrate that the bulk SOC can induce sizable spin-memory loss at a FM/HM interface. A remark is in order. Our mechanism based on the bulk SOC may look similar to other mechanisms based on the interfacial SOC in the sense that the bulk SOC can give rise to an interfacial SOC when it is combined with the broken inversion symmetry ($H_{\rm ISB}$ [Eq.~\eqref{eq:H_ISB}]) near the FM/HM interface. However as demonstrated below, the spin-memory loss by the bulk SOC can be significant even when the interfacial SOC arising from the bulk SOC and the inversion symmetry breaking is weak, $\sim$ 0.05 eV$\cdot$\AA. This is in contrast to the interfacial SOC based mechanisms of the spin memory loss, which require stronger interfacial SOC ($\sim$ 1 eV$\cdot$\AA) for the significant spin memory loss. A key element of the bulk SOC based mechanism is the spin-orbital entanglement. Since an electron incident on a FM/HM interface from the HM has its spin entangled to its orbital degree of freedom due to the bulk SOC, even "boring" orbital scattering at the interface can flip the electron spin and result in the spin-memory loss (see Fig.~\ref{fig:scattering} for illustration).
	
	%%%%%         Model         %%%%%
	
\begin{figure}[t!]
		\subfigure{\label{fig:fig1a}}
		\subfigure{\label{fig:fig1b}} 
		\subfigure{\label{fig:fig1c}}
		\subfigure{\label{fig:fig1d}} 
		\subfigure{\label{fig:fig1e}} 
		\begin{center}
			\includegraphics[width=8.5cm]{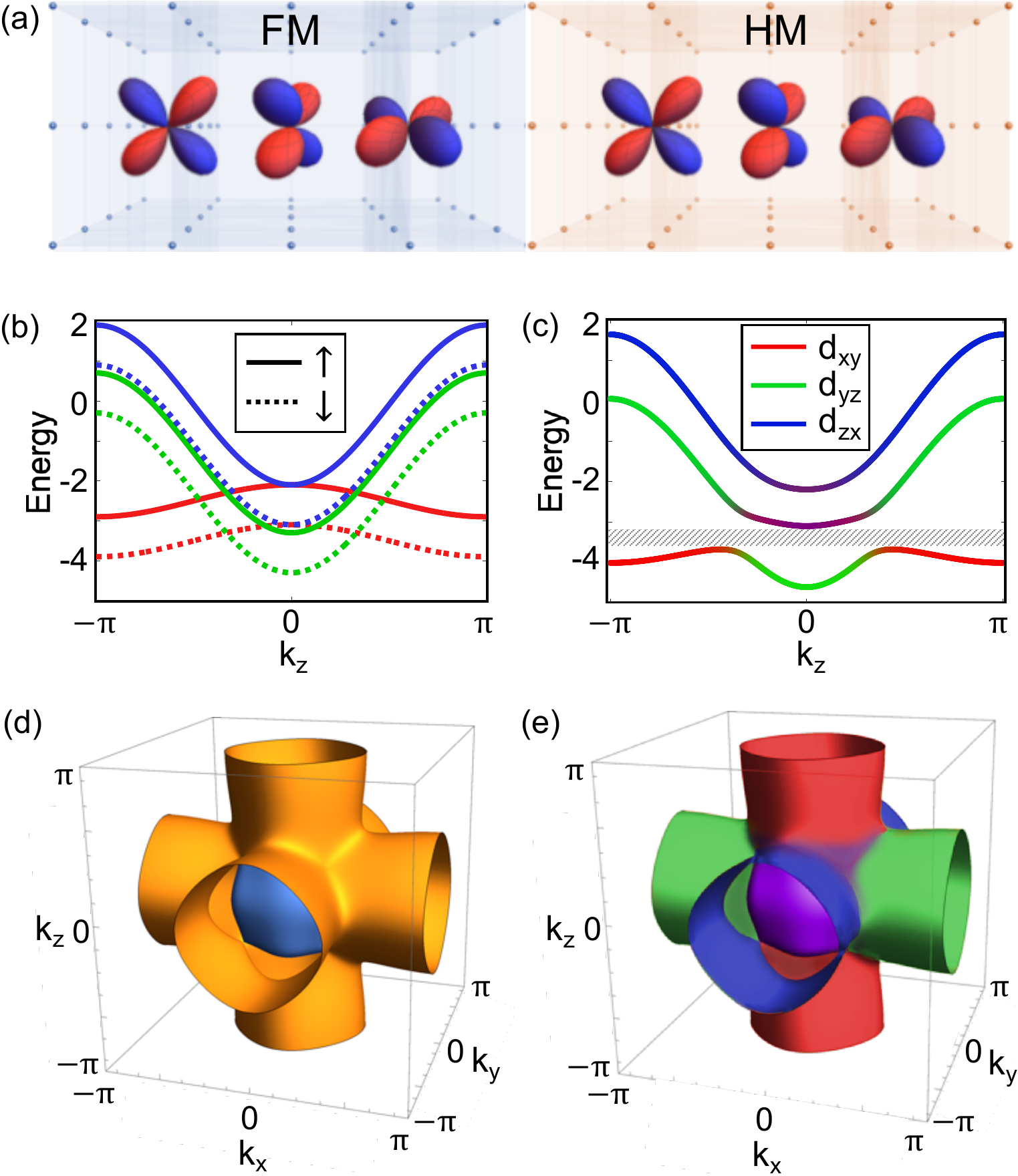}
		\end{center}
		\caption[short caption]{(a) Schematic figure of a FM/HM bilayer with both FM and HM in the simple cubic lattice.
			Each lattice site can host $t_{2g}$ $d$-orbitals, $d_{yz}$, $d_{zx}$ and $d_{xy}$.
			(b),(c) Energy band structures of a FM (b) and a HM (c) for $(k_x, k_y)=(\frac{\pi}{3}, 0)$. 
			The solid and dotted lines in (b) denote  the spin up and down bands, respectively. 
			In (b) and (c), the $d_{xy}$, $d_{yz}$ and $d_{zx}$ character bands are colored red, green, and blue, respectively. The gray dashed region in (c) denotes the avoided band crossing, near which the orbital character of the bands varies rapidly with ${\bf k}$.
			(d),(e) Three-dimensional plots of the Fermi surfaces in a HM at the Fermi energy $E_{\rm F}=-3.0$ eV. In (d), the colors are chosen to visualize the inner (blue) and outer Fermi surfaces. In (e), the colors represent the orbital character as in (c). Note that the orbital character varies rapidly with ${\bf k}$ near the avoided band crossing.
			The following parameter values are used throughout this paper; ($t_{\sigma}$, $t_{\pi}$, $J$, $\lambda$) = (1, 0.2, 0.5, 0.5)[eV].
		}
		\label{fig:fig1}
	\end{figure}

	A FM/HM bilayer is modeled as in Fig.~\ref{fig:fig1a}, where both FM ($z=-1,-2,\cdots$) and HM ($z=0,1,2,\cdots$) have the simple cubic lattice structure. Each lattice site can host $t_{2g}$ $d$-orbitals ($d_{xy}$, $d_{yz}$, and $d_{zx}$). The bilayer is described by the tight-binding Hamiltonian,  
	\begin{equation}
		H_\textup{tot}
		=
		H_{t_{2g}}+H_\textup{SOC}\Theta(z+0.5)+H_\textup{ex}\Theta(-z-0.5)+H_{\textup{ISB}},
		\label{eq:H_tot}
	\end{equation}
	where $H_{t_{2g}}$ is the kinetic energy term that describes the nearest-neighbor hoppings between the $t_{2g}$ orbitals. Since nearest-neighbor inter-orbital hoppings violate the inversion symmetry of the simple cubic lattice, $H_{t_{2g}}$ contains only intra-orbital hoppings  (for details, see supplementary material S1). We adopt the same intra-orbital hopping parameters for both FM and HM so that the inversion symmetry breaking is minimized at the FM/HM interface (thereby suppressing the interfacial SOC unless $H_{\rm ISB}$ is turned on). The FM and HM are distinguished only by the bulk SOC $H_{\rm SOC}$ for the HM and the exchange coupling $H_{\rm ex}$ for the FM. $\Theta(z)$ in Eq.~\eqref{eq:H_tot} is the Heaviside step function. $H_{\rm SOC}$ for the HM ($z\ge0)$ reads
	\begin{equation}
		H_\textup{SOC}
		=
		\frac{2 \lambda}{\hbar^2} \sum_{i,\sigma,\sigma'}C_{i,\sigma}^+\mathbf{L} \cdot \mathbf{S} C_{i,\sigma'},
		\label{eq:H_SOC}
	\end{equation}
	and $H_{\rm ex}$ for the FM ($z<0$) reads
	\begin{equation}
		H_\textup{ex}
		=
		J\sum_{i,\sigma,\sigma'}C_{i,\sigma}^+ \ \mathbf{S} \cdot \mathbf{M} \ C_{i,\sigma'},
	\end{equation}
	where 
	$C^+_{i,\sigma}=(c^+_{i,xy,\sigma},c^+_{i,yz,\sigma},c^+_{i,zx,\sigma})$ is the three-component electron creation operator at the lattice site $i$ with spin $\mathbf{\sigma}$.
	$\mathbf{L}$ and $\mathbf{S}$ are the orbital and spin angular momenta of the $t_{2g} \ d$-orbitals, and $\mathbf{M}$ is a three-dimensional unit vector of magnetization.
	Here $\lambda$ and $J$ are the atomic SOC strength and the exchange interaction energy, respectively.

	To assess effects of the inversion symmetry breaking, we introduce $H_{\rm ISB}$, 
	\begin{equation}
		\begin{split}
			H_{\textup{ISB}} = \gamma \sum\limits_{i,\sigma} [C_{i,xy,\sigma}^\dagger C_{i+\hat{x},yz,\sigma} + C_{i,xy,\sigma}^\dagger C_{i+\hat{y},zx,\sigma} + \textup{H.c.}]
			\\
			-\gamma \sum\limits_{i,\sigma} [C_{i,xy,\sigma}^\dagger C_{i-\hat{x},yz,\sigma} + C_{i,xy,\sigma}^\dagger C_{i-\hat{y},zx,\sigma} + \textup{H.c.}],
		\end{split}
		\label{eq:H_ISB}
	\end{equation}
	where the sum over the lattice index $i$ runs only within the $z=0$ (interfacial layer of the HM) and the $z=-1$ (interfacial layer of the FM) layers since the inversion symmetry breaking is strongly localized in the two layers. $H_{\rm ISB}$ describes the nearest-neighbor {\it inter-orbital} hoppings between the $d_{xy}$ orbital and the $d_{yz}$ ($d_{zx}$) orbital along the $x$ ($y$) direction. These inter-orbital hoppings can exist only when the inversion symmetry is broken~\cite{Petersen2000}. Thus the inter-orbital hopping strength $\gamma$ in $H_{\rm ISB}$ can be regarded as the strength of the inversion symmetry breaking. As demonstrated in Ref.~\cite{Park2013}, $H_{\textup{ISB}}$ introduces the orbital-momentum coupling (see supplementary material S1), which, near the $\Gamma$ point, is proportional to $\mathbf{L} \cdot (\mathbf{k} \times \hat{\bf z})$. Combined with $H_{\rm SOC}$, $H_{\rm ISB}$ generates the Rashba spin-momentum coupling~\cite{Petersen2000,Park2013}.

	%%%%%        Result        %%%%%
	
	Eigenstates of the HM are two-fold degenerate [Fig.~\ref{fig:fig1}(c)] due to the inversion symmetry whereas the degeneracy is lifted in the FM [Fig.~\ref{fig:fig1}(b)] due to $H_{\rm ex}$. Scattering eigenstates (Fig.~\ref{fig:scattering}) of the FM/HM bilayer are  calculated by matching the incident, transmitted, and reflected waves through the scattering boundary conditions  (see supplementary material S2 for details). The matching also takes into account evanescent waves localized near the FM/HM interface.  For concreteness, ${\bf M}=\hat{\bf z}$ is assumed, which motivates the $z$ axis to be used as the spin quantization axis, although spin-up ($\sigma=\uparrow$) states are inevitably superposed with spin-down ($\sigma=\downarrow$) states due to $H_{\rm SOC}$ in the HM. Figure~\ref{fig:scattering} illustrates the nature of the superposition between states with different spins.
	
	To examine the spin-flip at the FM/HM interface, we calculate the spin currents carried by incident, reflected, and transmitted waves, separately. The conventional definition of the spin current is used, $i_{s_z,X} = \bra{\psi_X}\frac{1}{2}\{\sigma_z, \rm{v}_z\}\ket{\psi_X}$, where $\sigma_z$ and $\rm{v}_z$ are the z-directional spin and velocity. Here $|\psi_X\rangle$ denotes incident ($X=I$), reflected ($X=R$), or transmitted ($X=T$) wave. The $z$-polarized spin current is calculated since the spin $z$ component flip is entirely due to the spin-memory loss whereas the spin $x$ or $y$ components can be flipped through $H_{\rm ex}$ even without the spin-memory loss. We also calculate the charge current $i_{ch,X}$ carried by incident ($X=I$), reflected ($X=R$), and transmitted ($X=T$) waves, separately. For each scattering state, the spin-memory retention rate $\Delta_{R(T)}$ during the reflection ($R$) and transmission ($T$) is defined by
	\begin{equation}
		\Delta_{R(T)}
		=
		%\frac{\Delta i_{s_z,R/T}}{\Delta i_{ch,R/T}}.
		\frac{i_{s_z,R(T)}/i_{s_z,I}}{i_{ch,R(T)
			}/i_{ch,I}}.
		\label{eq:spin-memory-retention}
	\end{equation}
	Note that Eq.~\eqref{eq:spin-memory-retention} is defined so that pure charge scattering does not suppress $\Delta_{R(T)}$ (see supplementary material S3). Thus if $\Delta_{R(T)}$ is smaller than 1, it implies the spin-memory loss. We also define the spin-flip probability $P_{R(T)}$ for the reflection and transmission, 
	\begin{equation}
		P_{R(T)} = ( \frac{1 - \Delta_{R(T)}}{2}  ) \cross 100 \ [\%].
		\label{eq:spin-flip-prob}
	\end{equation}

	\begin{figure}[t!]
		\subfigure{\label{fig:calc_HF_R}}
		\subfigure{\label{fig:calc_HF_T}}
		\begin{center}
			\includegraphics[width=8.5cm]{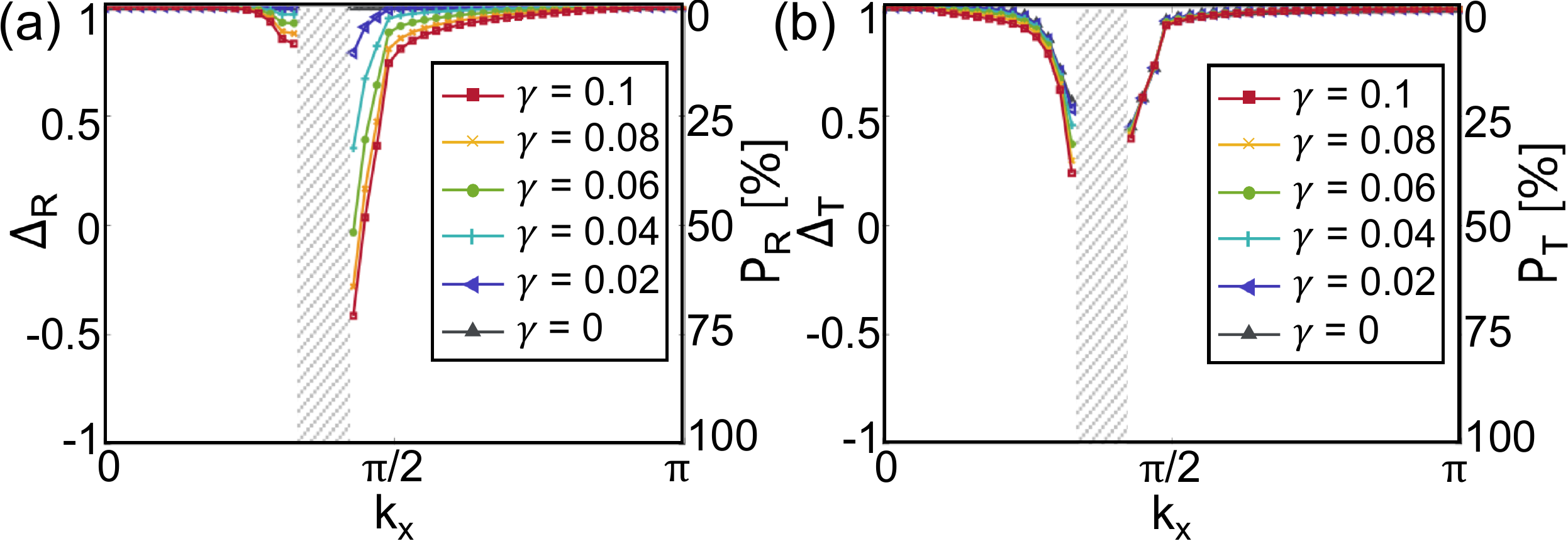}
		\end{center}
		\caption{
			The spin-memory retention rate $\Delta_{R(T)}$ and the spin-flip probability $P_{R(T)}$ for the reflection (a) and the transmission (b) at the Fermi energy $E_{\rm F}=-3.0$ eV as a function of the in-plane momentum $k_x$. The other component $k_y$ of the in-plane momentum is set to zero. Results for different strength $\gamma$ of $H_{\rm ISB}$ are denoted by different colors. The gray dashed region corresponds to  the avoided band crossing between the orange- and blue-colored bands in Fig.~\ref{fig:fig1d}.
			%The change in spin information and spin-flip probability for different ISB strengths for the spin current flowing from HM to FM.
			%The incident state is spin up-dominated state of a HM.
			%The gray dashed region is the band gap of the HM [Fig.~\ref{fig:fig1c}].
			%Here $k_y$ is fixed to be 0 and $k_x$ ranges from 0 to $\pi$ with $\lambda$ of 0.5eV.
		}
		\label{fig:calc}
	\end{figure}
	
	Figure~\ref{fig:calc} shows $\Delta_{R(T)}$ and $P_{R(T)}$ at the Fermi energy $E_{\rm F}=-3.0$ eV as a function of $k_x$ while $k_y$ is fixed to zero. For the reflection [Fig.~\ref{fig:calc_HF_R}], $P_R$ vanishes when the strength $\gamma$ of $H_{\rm ISB}$ is zero, and increases as $\gamma$ increases. But the increase of $P_R$ is not uniform over the Fermi surface but instead strongly localized near the gray dashed region, where the avoided crossing between energy bands of different orbital character occurs [Figs.~\ref{fig:fig1c}, \ref{fig:fig1d}] and thus orbital character of the Fermi surface varies rapidly with ${\bf k}$ [Fig.~\ref{fig:fig1e}]. Similarly, the spin-flip probability $P_T$ for the transmission [Fig.~\ref{fig:calc_HF_T}] also shows large values only near the gray dashed region. Interestingly, $P_T$ depends on $\gamma$ weakly and can have sizable values even when $\gamma$ vanishes. Reasons for the enhanced $P_{R(T)}$ are discussed below. As a side remark, we mention that $P_R$ may go above 50\% near the gray dashed region for $\gamma=0.1$ eV$\cdot$\AA\,, implying that the spin-memory is not merely lost but instead {\it reversed}.
	
	%%%%%  Theoretical analysis  - Mechanism  %%%%%
	
	To understand the origin of the spin-memory loss, we first examine the effect of $H_{\rm SOC}$ on the wave function structure of eigenstates in the HM. $H_{\rm SOC}$ is proportional to
	\begin{equation}
		\mathbf{L} \cdot \mathbf{S} = L_z S_z + \frac{1}{2} (L_- S_+ + L_+ S_-),
	\end{equation}
	where $L_z$ and $S_z$ are the z-directional orbital and spin angular momenta. The two last terms $L_-S_+$ and $L_+S_-$ convert $|d_{zx},\sigma\rangle$ and $|d_{yz},\sigma\rangle$ to $|d_{xy},-\sigma\rangle$, where $\sigma$ denotes $\uparrow$ or $\downarrow$. Thus both of the two-fold degenerate eigenstates, $|\phi_{\rm HM,1}\rangle$ and $|\phi_{\rm HM,2}\rangle$, are superpositions of spin up and down components,
	\begin{equation}
		\begin{split}
			\ket{\phi_\textup{HM,1}} = c_{yz}^{u} \ket{d_{yz}, \uparrow} + c_{zx}^{u} \ket{d_{zx}, \uparrow} + c_{xy}^{d} \ket{d_{xy}, \downarrow},
			\\
			\ket{\phi_\textup{HM,2}} = c_{yz}^{d}\ket{d_{yz}, \downarrow}+ c_{zx}^{d} \ket{d_{zx}, \downarrow}+ c_{xy}^{u}\ket{d_{xy}, \uparrow}.
		\end{split}
		\label{eq:HM-wavefunctions}
	\end{equation}
	Note that for each of $|\phi_{\rm HM,1}\rangle$ and $|\phi_{\rm HM,2}\rangle$, the spin up and down components have different orbital characters. Thus $|\phi_{\rm HM,1}\rangle$ and $|\phi_{\rm HM,2}\rangle$ are spin-orbital entangled states. The degree of the entanglement is weak far away from the avoided crossing [Fig.~\ref{fig:fig1d}] but the entanglement becomes strong near the avoided crossing.

	\begin{figure}[t!]
		\begin{center}
			\includegraphics[width=8.3cm]{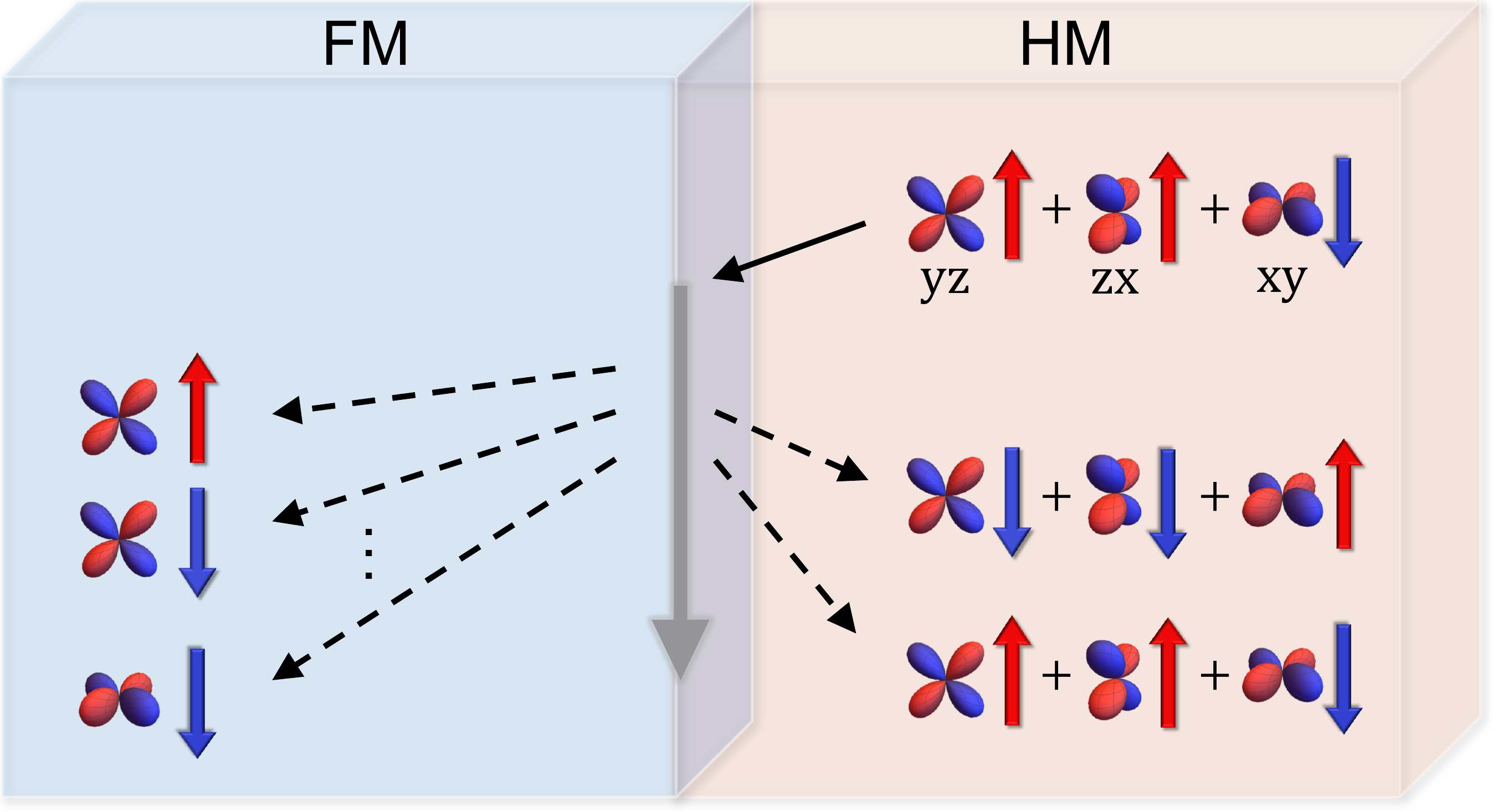}
		\end{center}
		\caption
		{
			Schematic figure of the scattering process in a FM/HM heterostructure. An incident wave (black solid arrow) is scattered at the FM/HM interface to generate outgoing waves (dashed black arrows). The gray vertical arrow at the interface denotes the evanescent wave. 
			The incident and reflected waves in the HM are inevitably superpositions of spin up (red up arrows) and down (blue down arrows) components due to the spin-orbital entanglement by $H_{\rm SOC}$. Orbital characters of the incident, reflected, and transmitted waves are schematically depicted. 
		}
		\label{fig:scattering}
	\end{figure}

	The spin-orbital entanglement plays an important role for the spin-memory loss. Figure~\ref{fig:scattering} shows schematically the scattering process for an incident wave $\ket{\psi_{I}}$ with wave function character $\ket{\phi_\textup{HM,1}}$. Suppose $c_{yz}^u$ or $c_{zx}^u$ is larger than $c_{xy}^d$, so that $\ket{\psi_{I}}$ carries a spin current with the spin polarization up. When $H_{\rm ISB}$ is absent ($\gamma=0$), the spin polarization of the transmitted wave arising from the incident wave is determined by how easily the three components $\ket{d_{yz},\uparrow}$, $\ket{d_{zx},\uparrow}$, $\ket{d_{xy},\downarrow}$ in the incident wave match the wave function character in the FM. Thus if the wave function in the FM has strong $\ket{d_{xy},\downarrow}$ character at $E_{\rm F}$, the spin polarization of the transmitted wave $\ket{\psi_T}$ has sizable spin down component, implying that mostly spin up nature of the incident wave is suppressed. Figure~\ref{fig:fig1b} indeed shows sizable $\ket{d_{xy},\downarrow}$ character at $E_{\rm F}=-3.0$ eV. This explains the spin-memory loss in $P_T$ even for $\gamma=0$ [Fig.~\ref{fig:calc_HF_T}].
	
	Next, we turn on $\gamma$ and investigate the effect of $H_{\rm ISB}$ on the spin-memory loss during the reflection ($P_R$). For small $\gamma$, the effect of $H_{\rm ISB}$ on $P_R$ can be analyzed perturbatively through  $\bra{\psi_{R}} H_\textup{ISB} \ket{\psi_{I}}$ with $\ket{\psi_I}$ of $\ket{\phi_{\rm HM,1}}$ character and $\ket{\psi_R}$ of $\ket{\phi_{\rm HM,2}}$ character. Here $\ket{\phi_{\rm HM,1}}$ and $\ket{\phi_{\rm HM,2}}$ carry the the spin currents with opposite spin polarization, and thus the magnitude of $\bra{\psi_{R}} H_\textup{ISB} \ket{\psi_{I}}$ can be used as a measure of the spin-flip through $H_{\rm ISB}$. According to Eq.~\eqref{eq:H_ISB}, $H_{\rm ISB}$ converts $\ket{d_{zx},\sigma}$ and $\ket{d_{yz},\sigma}$ to $\ket{d_{xy},\sigma}$ (or vice versa).  This change in the orbital character makes $\bra{\psi_{R}} H_\textup{ISB} \ket{\psi_{I}}$ nonvanishing, resulting in the spin-flip. Here we emphasize that $H_{\rm ISB}$ [Eq.~\eqref{eq:H_ISB}] does not modify the spin degree of freedom at all but the spin is effectively flipped nevertheless since the spin and orbital degrees of freedom are entangle in $\ket{\phi_{\rm HM,1}}$ and $\ket{\phi_{\rm HM,2}}$ [Eq.~\eqref{eq:HM-wavefunctions}]. Together with the fact that the spin-orbital entanglement becomes strong near the avoided crossing, this explains why $P_R$ increases with $\gamma$ especially near the avoided crossing [Fig.~\ref{fig:calc_HF_R}].
	
	%%%%%      Discussion      %%%%%
	
	It is worth comparing the present mechanism based on the bulk SOC with those~\cite{Chen2015a,Chen2015,Amin2016,Borge2017} based on the interfacial Rashba spin-momentum coupling. There are connections and also differences. The bulk SOC mechanism is connected to the Rashba spin-momentum coupling in the sense that $H_{\rm ISB}$ and $H_{\rm SOC}$, which are the core elements of the spin-memory loss for the reflection $P_R$, give rise to the Rashba spin-momentum coupling as well~\cite{Petersen2000,Park2013}. This connection can be understood by recalling the recognition~\cite{Park2013} that $H_{\rm ISB}$ amounts to the orbital-momentum coupling (see supplementary material S1), which has a particularly simple form $\sim {\bf L}\times {\bf k}\cdot \hat{\bf z}$ near the $\Gamma$ point. When combined with $H_{\rm SOC}\sim {\bf S}\cdot{\bf L}$, $H_{\rm ISB}$ generates the Rashba spin-momentum coupling $\alpha_{\rm R} {\bf S}\times {\bf k}\cdot \hat{\bf z}$. It has been argued~\cite{Park2011} that the combination of $H_{\rm ISB}$ and $H_{\rm SOC}$ is an important source of the Rashba spin-momentum coupling in many systems. However the bulk SOC mechanism differs from the Rashba spin-momentum coupling mechanisms in that the latter mechanisms~\cite{Chen2015a,Chen2015,Amin2016,Borge2017} neglect the atomic orbital degree of freedom completely and thus the spin-orbital entanglement cannot play any role. We demonstrated above that the spin-orbital entanglement is responsible for the spin-memory loss during the transmission ($P_T$) even when $H_{\rm ISB}=0$ (thus Rashba spin-momentum coupling is absent). Also for the spin-memory loss during the reflection ($P_R$), for which $H_{\rm ISB}$ is essential (thus Rashba spin-momentum coupling is present), the spin-orbital entanglement allows for the sizable spin-memory loss even from weak $H_{\rm ISB}$ with $\gamma\sim$ 0.1 eV$\cdot$\AA, which amounts to the Rashba spin-momentum coupling strength $\alpha_{\rm R}$ of $\sim$ 0.05 eV$\cdot$\AA\, (see supplementary material S4 for the evaluation of $\alpha_{\rm R}$). This value is comparable to $\alpha_{\rm R}=0.03$ eV$\cdot$\AA\ for the weak Rashba system, Ag(111) surface~\cite{Ast2007}, but much smaller than the value $\sim 1$ eV$\cdot$\AA\, assumed in the theoretical studies~\cite{Chen2015a,Chen2015} of the spin-memory loss by the Rashba spin-momentum coupling.
	
	An important implication of the bulk SOC mechanism is that the degree of the spin-memory loss is not uniform in the momentum space but is strong near the band crossing (Fig.~\ref{fig:calc}) because the spin-orbital entanglement is strong there. It is interesting that the first-principles calculations of spin-flip scattering probabilities for Cu/Pd interfaces~\cite{Belashchenko2016} also find large spin-flip probabilities near band crossing points, where the orbital hybridization occurs and the spin-orbital entanglement becomes strong. Although our result (Fig.~\ref{fig:calc}) based on the simple Hamiltonian $H_{\rm tot}$ has clear limitations in terms of quantitative predictions, this qualitative agreement with the first-principles calculation result suggests that the bulk SOC mechanism may be relevant in real materials. Finally we mention that the intrinsic spin Hall effect in HMs, which is an important mechanism of the spin current generation in FM/HM bilayers, arises mainly in the momentum space regions close to the band crossing where the orbital hybridization occurs~\cite{Guo2008, Tanaka2008, Go2018}. Thus the bulk SOC mechanism may be especially relevant for the spin-Hall-effect-induced spin transport.
	
	%%%%%       Summary        %%%%%
	
	To conclude, we investigated the spin-flip scattering at a FM/HM interface induced by the bulk SOC ($H_{\rm SOC}$) with explicit account of the atomic orbital degree of freedom. Even when the inversion symmetry breaking ($H_{\rm ISB}$) at the FM/HM interface is strongly suppressed, the spin-flip can still occur at the interface due to the spin-orbital entanglement caused by the bulk SOC. Additional spin-flip arises when the inversion symmetry breaking ($H_{\rm ISB}$) at the FM/HM interface is turned on. Our work proposes another spin-memory loss mechanism, in which the spin-orbital entanglement by the bulk SOC plays key roles.
	
	% \newcommand{\beginsupplement}{%
	% \setcounter{table}{0}
	% \renewcommand{\thetable}{S\arabic{table}}%
	% \setcounter{figure}{0}
	% \renewcommand{\thefigure}{S\arabic{figure}}%
	%  }
	
	%\section*{Supplementary Material}
	\bigskip
	See the supplementary material for more details on the tight-binding model for a FM/HM structure, scattering boundary conditions, and additional information on the evaluation of the Rashba spin-momencum coupling.
	
	\bigskip
	%\section*{Acknowledgments}
	% This work was supported by the Samsung Science \& Technology Foundation (Grant Nos.BA-1501-07 \& BA-1501-57). We acknowledge J. Sohn, D. Go, and S. Cheon for fruitful discussions.
	%\begin{acknowledgments}
	This work was supported by the Samsung Science \& Technology Foundation (Grant Nos.BA-1501-07 \& BA-1501-51). We thank J. Sohn, D. Go, and S. Cheon for fruitful discussions.
	%\end{acknowledgments}
	
	\section*{Data Availability}
	The data that support the findings of this study are available from the corresponding author upon reasonable request.
	
	\nocite{*}
	\bibliography{SML_ML_HWL}% Produces the bibliography via BibTeX.
	
\end{document}